\documentclass[12pt]{iopart}

\usepackage{graphicx}
\usepackage{dcolumn}
\usepackage{bm}
\usepackage{hyperref}
\usepackage{acronym}
\usepackage{color}
\usepackage{subfigure}
\usepackage{multirow}
\usepackage{amssymb,amsfonts}%
\usepackage{amsthm}%
\usepackage{orcidlink}

\newacro{GW}{gravitational wave}
\newacro{BBH}{binary black hole}
\newacro{MBHB}{massive black hole binary}
\acrodefplural{MBHB}[MBHBs]{massive black hole binaries}
\newacro{SBBH}{stellar mass binary black hole}
\newacro{SNR}{signal-to-noise ratio}
\newacro{LIGO}{Laser Interferometer GW Observatory}
\newacro{LISA}{Laser Interferometer Space Antenna}
\newacro{PTA}{pulsar timing array}
\newacro{TDI}{time delay interferometry}
\newacro{PSD}{power spectral density}
\newacro{ASA}{All-Sky Average}
\newacro{LFA}{Low-Frequency Approximation}
\newacro{SWA}{Short-Wave Approximation}

\begin{document}

\title[SNR Analytic Formulae of the Inspiral BBHs in TQ]{Signal-to-noise Ratio Analytic Formulae of the Inspiral Binary Black Holes in TianQin}

\author{Hong-Yu Chen$^1$\orcidlink{0009-0006-2843-7409}, 
        Han Wang$^1$\orcidlink{0009-0007-5095-9227}, 
        En-Kun Li$^{1,*}$\orcidlink{0000-0002-3186-8721}, 
        Yi-Ming Hu$^{1,*}$\orcidlink{0000-0002-7869-0174}}
\address{$^1$MOE Key Laboratory of TianQin Mission, %
            TianQin Research Center for Gravitational Physics $\&$ School of Physics and Astronomy, %
            Frontiers Science Center for TianQin, %
            Gravitational Wave Research Center of CNSA, %
            Sun Yat-sen University (Zhuhai Campus), %
            Zhuhai, 519082, China}
\ead{lienk@sysu.edu.cn(En-Kun Li, Corresponding Author), huyiming@sysu.edu.cn(Yi-Ming Hu, Corresponding Author)}

\begin{abstract}
Binary black holes are one of the important sources for the TianQin gravitational wave project. 
Our research has revealed that, for TianQin, the signal-to-noise ratio of inspiral binary black holes can be computed analytically. 
This finding will help to quickly estimate the detectability of gravitational waves from different types of binary black holes.
In addition, it provides us with the accumulation pattern of signal-to-noise ratios for binary black hole signals, which can be used for the identification of future signals and the construction of detection statistics.
In this paper, we demonstrated the signal-to-noise ratio analytic formulae from stellar-mass black holes to massive black holes.
With the all-sky average condition, the signal-to-noise ratio for most binary black hole signals can be determined with a relative error of $\lesssim10\%$, with notable deviations only for chirp masses near $1000~M_\odot$.
Moreover, the signal-to-noise ratio without the average includes only an additional coefficient, which we refer to as the response factor. Although this term is not easily calculated analytically, we provide a straightforward calculation method with an error margin of $1\sigma$ within 2\%.
\end{abstract}

\maketitle

\section{Introduction}

In recent years, the domain of \ac{GW} detection has undergone remarkable advancements, marking a significant milestone in the scientific community.
One of the key breakthroughs came in 2015, when \ac{LIGO} \cite{LIGO:2015CQGra} successfully detected the first \ac{GW} signal from a \ac{BBH} merger \cite{Abbott:2016PhRvD, Abbott:2016PhRvL}.
So far, hundreds of \ac{GW} detections have been made by ground-based detectors \cite{LIGOScientific:2018mvr, LIGOScientific:2020ibl, LIGOScientific:2021djp, GraceDB}. 
Furthermore, evidence for the stochastic \ac{GW} background has recently been reported by \ac{PTA} \cite{NANOGrav:2023gor, Antoniadis:2023ott, Reardon:2023gzh, Xu:2023wog}.
In parallel, research on space-based \ac{GW} detectors is also progressing \cite{Gong:2021NatAs}.
For example, the TianQin project \cite{Luo:2016CQGra}, a space-borne \ac{GW} detection mission, aims to detect low-frequency \acp{GW} and further advance our understanding of the universe \cite{Li:2024rnk}.

TianQin consists of three satellites positioned in a geostationary orbit approximately 100,000 kilometers above the Earth's surface, forming an equilateral triangle. 
These satellites emit and receive laser beams among themselves, effectively creating a space-based laser interferometer.
The principle of the TianQin detector lies in measuring minor variations in the distances between the satellites, which are caused by the passing of \acp{GW}. 
Influenced by the laser arm length and various instrumental noise factors of TianQin, the detector's sensitive frequency band ranges from 0.1 mHz to 1 Hz.
One of the main sources for TianQin is the \ac{BBH}. 
TianQin is expected to be capable of observing \ac{BBH} systems across cosmic history, from the early universe (with a redshift of approximately $z \sim 20$) to the present era \cite{Wang:2019PhRvD}. 
In this paper, we will discuss the \ac{MBHB} and \ac{SBBH} systems separately.

\Ac{MBHB}, as one of the most intense cosmic phenomena, is expected to accumulate sufficient \ac{SNR} on the TianQin detector during the inspiral phase, making it detectable before the merger. 
This capability presents us with a unique astrophysical laboratory to delve into the formation and evolution of black holes \cite{Sesana:2011PhRvD, Schnittman:2013CQGra}, trace the history of the cosmos \cite{Sesana:2011PhRvD, Zhu:2022PhRvR}, and test gravitational theories \cite{Berti:2015CQGra, Yagi:2016CQGra, Gair:2013LRR}.
Moreover, the early warnings of \ac{MBHB} systems before their merger could significantly advance the field of multi-messenger astronomy \cite{Chen:2023SCPMA}.
We have found that due to the unique sensitivity curve of TianQin, the \ac{SNR} of \acp{GW} emitted by inspiral \acp{MBHB} follows a special relationship:
\begin{eqnarray}
\rho^2 \propto \frac{\rm observation \ time}{\rm time \ to \ coalescence}
\end{eqnarray}

For \acp{SBBH}, they merge at high frequencies. Space-based \ac{GW} detectors can observe their early inspiral phase within the lower frequency range, with the SNR reaching over a dozen \cite{Liu2020, Buscicchio2021}.
The detection of \acp{SBBH} in the millihertz band with a long observation time enables more precise measurements of parameters such as mass and sky location, while also preserving the evolution of key properties such as eccentricity and spin \cite{Liu2020, Buscicchio2021}.
These capabilities not only help constrain the formation mechanisms of such systems \cite{Nishizawa2016, Wang2024}, but also facilitate measuring the Hubble parameter or constraining certain parameters in modified gravity theories with great precision \cite{Barausse:2016eii, Kyutoku:2016zxn, Chamberlain:2017fjl, DelPozzo:2017kme}.
We have found that, for TianQin, the \ac{SNR} of \acp{GW} emitted by inspiral \acp{SBBH} follows a special relationship:
\begin{eqnarray}
\rho^2 \propto \sqrt{{\rm observation \ time}}-\sqrt{{\rm time \ to \ coalescence}}
\end{eqnarray}

These relationships will significantly facilitate the estimation of the \ac{SNR} for the inspiral phase of \acp{BBH}, simplify the assessment of the detectability of different \ac{BBH} models, and be beneficial for the identification of future signals and the construction of detection statistics.
In this paper, we first present a general \ac{SNR} calculation method for the inspiral phase. Then, we derive the \ac{SNR} relationships for \ac{MBHB} and \ac{SBBH} using \ac{LFA} and \ac{SWA} approximations respectively. For \ac{MBHB}, we study the inspiral \ac{SNR} relationships under both \ac{ASA} (Equation \ref{equ:ASA_result}) and non-average (Equation \ref{equ:nonASA_result}) conditions. For \ac{SBBH}, we only focus on the inspiral \ac{SNR} relationship under \ac{ASA} (Equation \ref{equ:SBBH_result}).

The organization of this paper is as follows.
Section \ref{sec:SNR_calculation} will outline the formula for calculating the \ac{SNR}. 
Section \ref{sec:MBHB} explores the \ac{SNR} accumulation for \ac{MBHB} systems in the TianQin detector, under \ac{ASA} and non-average conditions respectively. 
Then, Section \ref{sec:SBBH} delves into the feasibility of formulating an analytical \ac{SNR} formula for the inspiral \acp{SBBH}.
We conclude with a summary in Section \ref{sec:conculsion}.

\section{SNR calculation}\label{sec:SNR_calculation}

The SNR can be described as the ratio of the signal amplitude to the sensitivity of the detector.
The optimal \ac{SNR} is commonly computed using the inner product approach \cite{Cutler:1994PhRvD, Moore:2015CQGra}:
\begin{eqnarray}
\rho = \sqrt{\langle h \mid h \rangle},
\end{eqnarray}
where $h$ is the waveform of the signal, and $\langle a \mid b \rangle$ represents the inner product between $a$ and $b$ expressed as follows:
\begin{eqnarray}
\langle a \mid b \rangle = 4 \Re \int_0^\infty \frac{\widetilde{a} \left(f\right) \cdot \widetilde{b}^{\ast} \left(f\right) }{S_{n} \left(f\right)} {\rm d}f,
\end{eqnarray}
where $S_{n} \left(f\right)$ is the one-sided noise \ac{PSD} and $\Re$ is the real component.
The instrument settings used for the TianQin/LISA PSD are detailed in \ref{sec:det_set}.

During the inspiral phase, \ac{BBH} signals can be accurately approximated using post-Newtonian formulae \cite{Blanchet:2014LRR}, under which the frequency can be expressed as a function of observation time:
\begin{eqnarray}\label{equ:ft_relation}
f(t) = \frac{1}{8\pi} \left(\frac{G\mathcal{M}}{c^3}\right)^{-5/8} \left(\frac{t_c-t}{5}\right)^{-3/8}.
\end{eqnarray}
Here, $G$ and $c$ denote the gravitational constant and the speed of light, 
$t_c$ and $\mathcal{M}$ correspond to the coalescence time and the detector-frame chirp mass.
Setting the time origin at the onset of the observation period, the upper and lower limits of the frequency of the \ac{SNR} integration are $f_{\rm max}=f\left(T_{\rm obs}\right)$ and $f_{\rm min}=f(0)$, respectively.
Moreover, we need to cutoff the data to reduce errors if the signal frequency exceeds the cutoff frequency of TianQin. A detailed description can be found in the \ref{sec:det_set}.

In the computation of the optimal \ac{SNR}, we solely utilize the real component of the inner product. 
As a result, the phase of the waveform has a negligible impact on the calculation of the \ac{SNR}, enabling us to ascertain the \ac{SNR} value purely based on the waveform's amplitude. 
The amplitude of \ac{GW} emitted by inspiral \ac{BBH} systems can be represented by a simple post-Newtonian formula \cite{Husa:2016PhRvD}:
\begin{eqnarray}
\mathcal{A}(f) = \sqrt{\frac{2}{3\pi^{1/3}}} c^{-3/2} \frac{1}{D_L} \left(G\mathcal{M}\right)^{5/6} f^{-7/6},
\end{eqnarray}
where $D_L$ represents the luminosity distance.
After the frequency band and amplitude are known, calculating the \ac{SNR} involves computing the response function and the noise \ac{PSD}. 
Next, we will specifically introduce the accumulation relationship of \ac{SNR} over time under both \ac{ASA} and non-average scenarios. 

\section{Massive Black Holes}\label{sec:MBHB}

\subsection{All-Sky Average}\label{sec:SNR_with_ASA}

By using the \ac{LFA}, we can provide the analytic formula of the \ac{SNR} for the TianQin detector's observation of inspiral \acp{MBHB}:
\begin{eqnarray}\label{equ:ASA_result}
\rho (T_{\rm obs}) &= \sqrt{\frac{15}{2048}} \frac{L}{D_L} \frac{c}{\sqrt{N_a t_c}} \times \sqrt{\frac{t_{obs}}{t_c-t_{obs}}}, \nonumber \\
    &\approx 185.2 \times \frac{1~{\rm Gpc}}{D_L} \sqrt{\frac{1~{\rm week}}{t_c}} \times \sqrt{\frac{T_{\rm obs}}{t_c-T_{\rm obs}}},
\end{eqnarray}
A detailed description of the calculation process can be found in the \ref{sec:cal_asa}.
It is worth noting that the formula presented here does not include the chirp mass parameter. 
This indicates that due to TianQin's frequency power-law sensitivity curve, in specific frequency bands, variations in chirp mass do not significantly affect the \ac{SNR} of \ac{MBHB} \ac{GW} signals. 
This observation aligns with findings in previous work \cite{Chen:2023SCPMA}.

If there is a gap in the data, the \ac{SNR} can be expressed as:
\begin{eqnarray}
\rho = \sqrt{\rho^2\left(T_{\rm obs}\right) - \rho^2\left(t_{e}\right) + \rho^2\left(t_{s}\right)},
\end{eqnarray}
where, $t_{s}$ and $t_{e}$ indicate the start and end times of the gap. More gaps can also be calculated simply using this similar method.

\begin{figure}[htbp]
  \centering
  \subfigure[w/o correction.]{\includegraphics[width=0.45\textwidth]{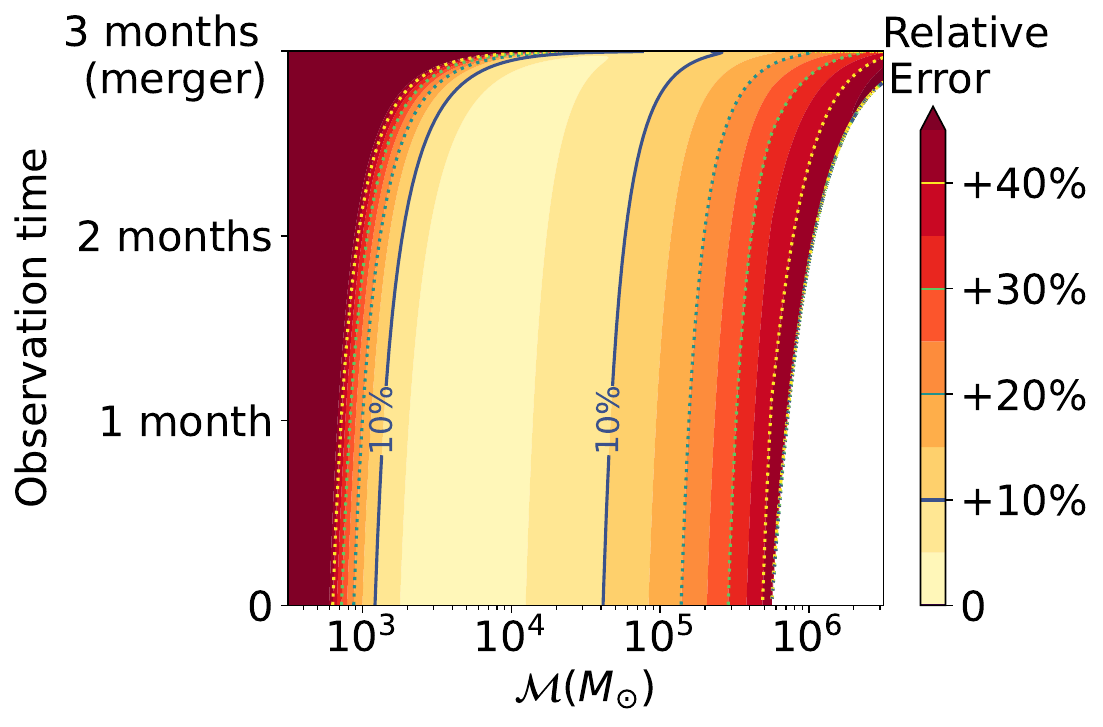}\label{fig:SNR_appro}}
  \subfigure[with correction.]{\includegraphics[width=0.45\textwidth]{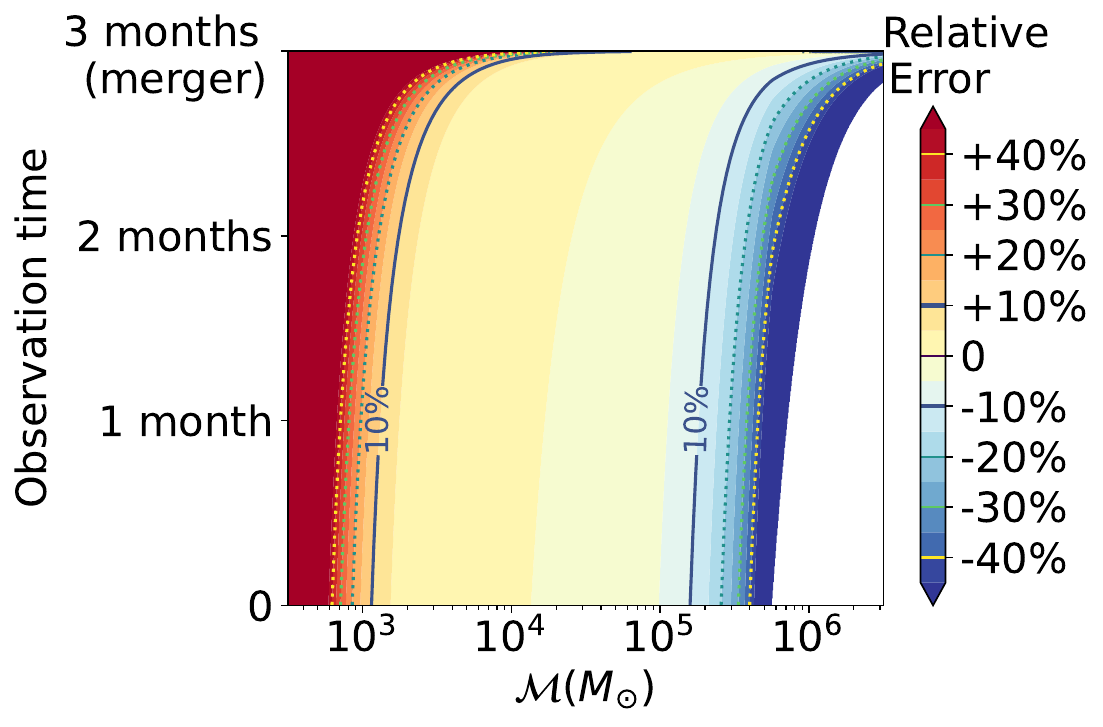}\label{fig:SNR_appro_1}}
  \caption{Estimated error of the \ac{SNR} for signals that merge at the third month, under different observation times and chirp mass conditions. The left and right panels respectively illustrate the estimated errors of Equation \ref{equ:ASA_result} and (\ref{equ:ASA_1o}). The blank area in the lower right corner indicates that the signal has not yet entered the sensitive frequency band of TianQin.}
  \label{fig:snr_ASA}
\end{figure}

To verify the precision of the estimation formula, we examined the error levels across different chirp masses and observation times.
In our test, the true signal is generated by the IMRPhenomD waveform \cite{IMRPhenomD1, IMRPhenomD2}, and all the \ac{BBH} systems are merged at the third month mark $\left(t_c=3~{\rm months}\right)$, as it illustrates the possible maximum observation time in the future. 
We define the relative error as
\begin{eqnarray}
\Delta = \frac{\rho - \rho_{\rm true}}{\rho_{\rm true}}. 
\end{eqnarray}
As depicted in Figure \ref{fig:SNR_appro}, for signals with chirp mass range from $10^3$ to $10^4 M_{\odot}$, the relative error is minimal. 
Signals outside this region will be overestimated by our formula.
Considering the low \ac{SNR} of the inspiral \acp{BBH}, the deviation caused by this relative error will be very low, ensuring the reliability of the formula.

Next, we consider the lowest-order correction to obtain more accurate results:
\begin{eqnarray}\label{equ:ASA_1o}
\rho _{\rm corr}^2 = \rho^2 + C. 
\end{eqnarray}
The details of the calculation method for the correction term are provided in the \ref{sec:cal_corr}. For TianQin, 
\begin{eqnarray}\label{equ:ASA_corr}
\fl C_{TQ} &= -\frac{3\pi c^{1/8}}{320000\times 5^{3/8}} \left(G\mathcal{M}\right)^{5/8} \left(\frac{L}{D_L}\right)^2 \frac{1}{N_a} \times \left[\left(t_c-t_{{\rm obs}}\right)^{-5/8}-t_c^{-5/8}\right], \nonumber \\
    \fl &\approx -1692 \left(\frac{\mathcal{M}}{10^4 ~M_{\odot}}\right)^{5/8} \left(\frac{1~{\rm Gpc}}{D_L}\right)^{2} \times \left[ \left(\frac{t_c-T_{\rm obs}}{1 ~{\rm week}}\right)^{-5/8}-\left(\frac{t_c}{1 ~{\rm week}}\right)^{-5/8} \right].
\end{eqnarray}
Figure \ref{fig:SNR_appro_1} illustrates the improvement brought by this correction. 
Compared to Figure \ref{fig:SNR_appro}, the corrected \ac{SNR} formula provides an accurate estimation for sources within the mass range of $2\times10^3 \sim 2\times10^5 M_\odot$.
Signals with masses falling below this range are prone to be overestimated by our formula, whereas signals with masses exceeding this range are susceptible to underestimation.
Systems with $\mathcal{M} \leq 10^3 M_\odot$ retain relatively poor estimation because the \ac{LFA} we used here is no longer suitable for these systems.

\begin{figure*}[htbp]
  \centering
  \subfigure[w/o correction.]{\includegraphics[width=0.45\textwidth]{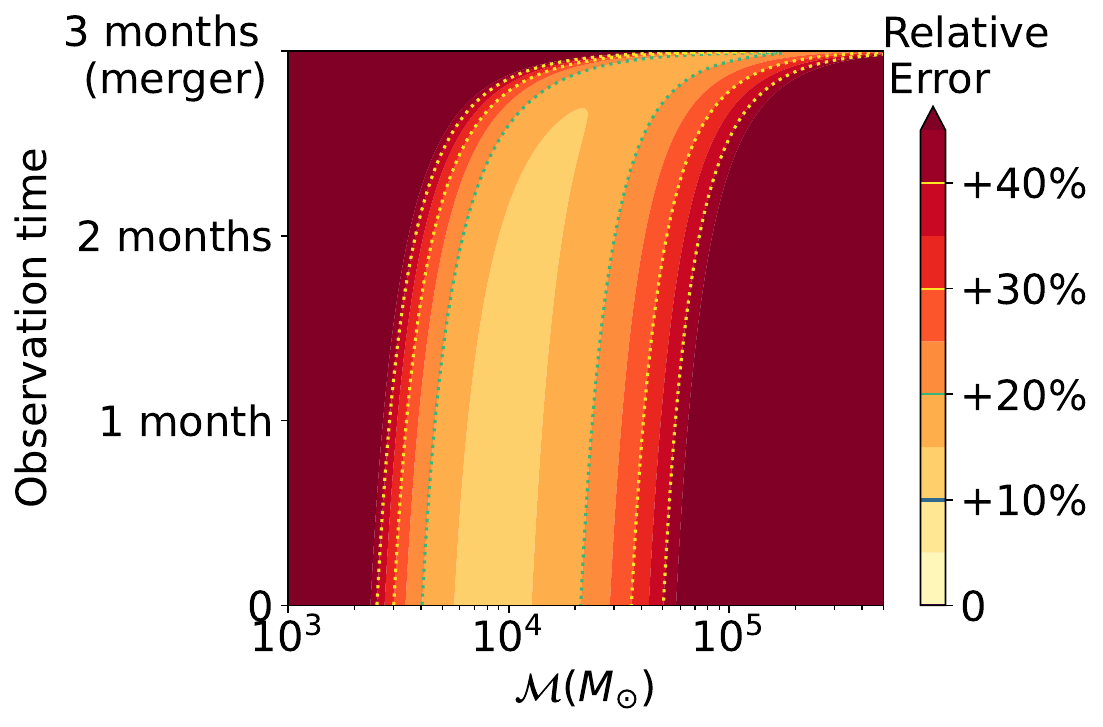}\label{fig:SNR_appro_LISA}}
  \subfigure[with correction.]{\includegraphics[width=0.45\textwidth]{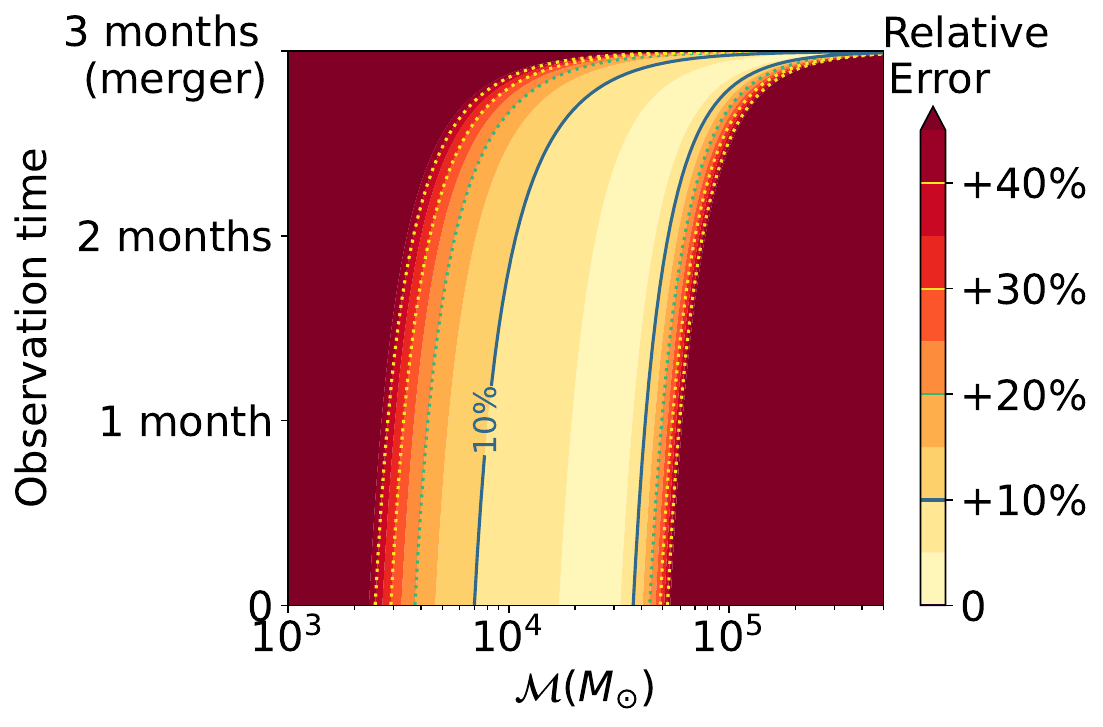}\label{fig:SNR_appro_LISA_1}}
  \caption{The \ac{SNR} estimation error for \ac{LISA} varies with the chirp mass and observation time. The left panel uses the simplest zeroth-order \ac{SNR} formula for estimation, while the right panel takes into account the lowest-order correction.}
  \label{fig:snr_ASA_LISA}
\end{figure*}

We've also noted that this formula might not be well-suited for other space-based \ac{GW} detectors, such as \ac{LISA} \cite{LISA:2017arXiv}.
\ac{LISA}, which will be placed in heliocentric orbit, consists of three spacecraft forming an equilateral triangle with each arm spanning 2.5 million kilometers. 
The \ac{ASA} sensitivity curve of \ac{LISA} aligns with that of TianQin, differing only in arm length and two noise terms, as detailed in \ref{sec:det_set}.
The increase in arm length enhances sensitivity in the low-frequency band, prompting us to select a lower cutoff frequency for \ac{LISA} at $10^{-5}$ Hz rather than $10^{-4}$ for TianQin.
Different dependencies of $S_a$ on frequency only alter the higher-order terms of the \ac{SNR} formula. 
For LISA, the lowest-order correction is:
\begin{eqnarray}\label{equ:LISA_1o}
\fl C_{LISA} &= -\frac{3\pi^2}{2\times10^6\times(5 c)^{7/4}} \left(G\mathcal{M}\right)^{5/4} \left(\frac{L}{D_L}\right)^2 \frac{1}{N_a} \times \left[\left(t_c-T_{\rm obs}\right)^{-1/4}-t_c^{-1/4}\right], \nonumber \\
    \fl &\approx -48298 \left(\frac{\mathcal{M}}{10^4~M_{\odot}}\right)^{5/4} \left(\frac{1~{\rm Gpc}}{D_L}\right)^2 \times \left[\left(\frac{t_c-T_{\rm obs}}{1~{\rm week}}\right)^{-1/4}-\left(\frac{t_c}{1~{\rm week}}\right)^{-1/4}\right]
\end{eqnarray}
However, as shown in Figure \ref{fig:snr_ASA_LISA}, the estimated results of the \ac{SNR} formula always have significant errors, regardless of whether the lowest-order correction is taken into account. 
Due to experiments like LISA Pathfinder, LISA has gained a more profound understanding of noise characteristics. Consequently, it has developed a more complex formula for the acceleration noise power spectrum. Applying our simplified approaches to LISA may introduce larger errors. However, such a simplification will not result in significant errors for TianQin with a relatively simpler noise design curve.
When not considering the lowest-order correction, as shown in Figure \ref{fig:SNR_appro_LISA}, the estimated error remains above 10\%. 
Even with the lowest-order correction, only sources with masses around $3\times 10^4 M_{\odot}$ can be estimated accurately.
Other LISA-like missions (e.g., Taiji) have similar orbits and noise power spectral density, so their \ac{SNR} formulas' structures and applicability are basically the same as LISA's.
Furthermore, we attempted to vary the cutoff frequencies for LISA, with the low-frequency cutoff ranging from $10^{-5}$ Hz to $10^{-4}$ Hz, and the high-frequency cutoff ranging from 0.1 Hz to 1 Hz.
Our findings indicate that there is no substantial difference in the magnitude of estimation errors.

\subsection{Non-average} \label{sec:SNR_with_non_ASA}

The \ac{ASA} result yields a mere approximation of expected values. 
The \ac{SNR} of \ac{GW} sources can vary greatly depending on their sky positions (latitude and longitude) and polar angles (polarization and inclination angle) in actual detection. 
We found that after considering the response, the estimated \ac{SNR} is only multiplied by a factor compared to the \ac{ASA} case, i.e.
\begin{eqnarray}\label{equ:nonASA_result}
\rho (T_{\rm obs}) = P_{\rm res} \times \rho_{\rm ASA} (T_{\rm obs}),
\end{eqnarray}
where the response factor $P_{\rm res}$ is a function of the sky position and polar angles, defined as:
\begin{eqnarray}
P_{\rm res}^2 = \frac{8\pi}{9} \left(P_{12}^2 + P_{23}^2 + P_{31}^2 - P_{12}P_{31} - P_{23}P_{31} - P_{12}P_{23}\right).
\end{eqnarray}
The expression $P_{ab} = n_l \cdot P^{22} \cdot n_l$ defines the inner product of the 2-2 mode polarization tensor $P^{22}$ and the link unit vectors $n_l$ between two satellites $a$ and $b$.
Note that this concept is distinct from the antenna pattern and focuses on the ratio of the response at a specific sky position and polar angle to the \ac{ASA}.
A detailed description of the calculation process can be found in the \ref{sec:cal_non_asa}.
And we offer a simple and quick calculation program in \url{https://github.com/TianQinSYSU/responseFactor}.

\begin{figure*}[htbp]
  \centering
  \subfigure[Response factor of sky position]{\includegraphics[width=0.45\textwidth]{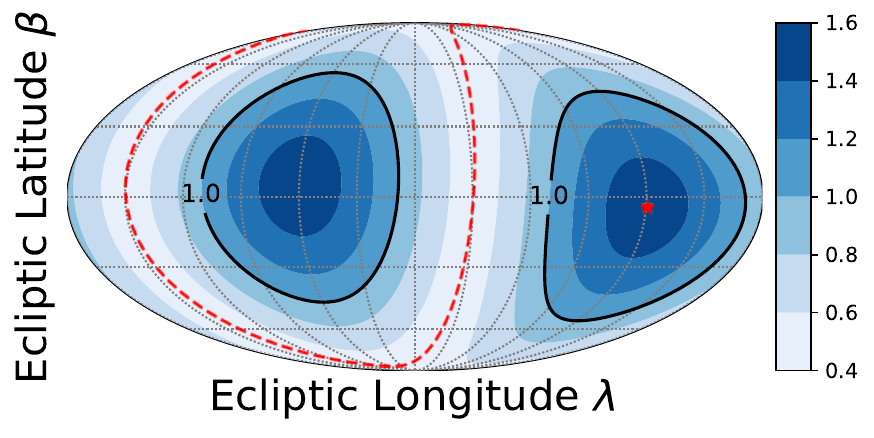}\label{fig:sky}}
  \subfigure[Response factor of polar angle]{\includegraphics[width=0.45\textwidth]{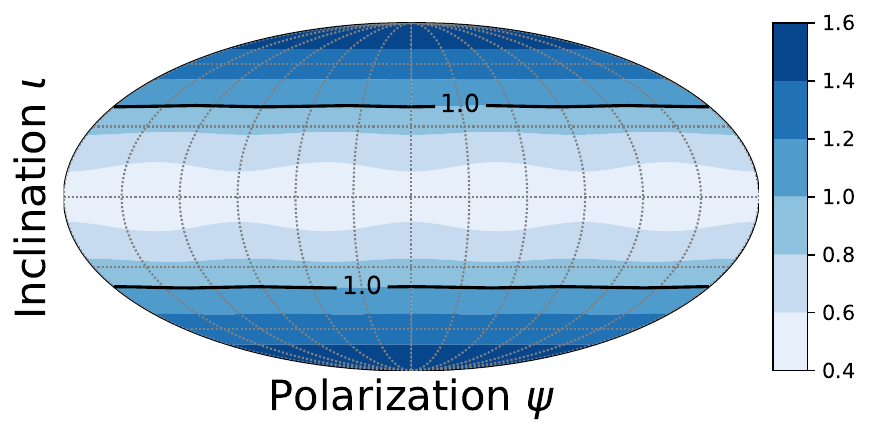}\label{fig:polar}}
  \caption{The left panel illustrates the variation of the response factor with respect to sky position, with the TianQin orbit plane marked by a red dashed line and the position of the double white dwarf system RX J0806.3 + 1527 marked by a red asterisk, signifying the direction in which the TianQin constellation is pointed. 
  The right panel depicts the variation of the response factor with respect to the polar angle. 
  Both figures highlight where the response factor equals $1.0$ with a black line.}
  \label{fig:ResponseFactor}
\end{figure*}

Figure \ref{fig:ResponseFactor} illustrates the variation of the response factor across different sky positions and polar angles.
Given the response factor's dependence on all four positional angles, we integrate the remaining two angles to ascertain their relationship with a pair of selected angles during the computation process.
As shown in Figure \ref{fig:sky}, the TianQin constellation exhibits pronounced sensitivity towards the double white dwarf system RX J0806.3 + 1527 (hereafter J0806) and the reciprocal direction, attributable to its orbital plane's always facing the J0806. 
The orbital plane of TianQin and the coordinates of J0806 in the ecliptic reference frame are marked using a red dashed line and a red asterisk, respectively.
In Figure \ref{fig:polar}, it is evident that the inclination angle $\iota$ has a more significant impact on the \ac{SNR} compared to the polarization angle $\psi$. The \ac{SNR} is maximized when the inclination angle approaches $0$ or $\pi$, indicating that the detector is face-on to the source.
Conversely, the \ac{SNR} is minimal when the inclination angle is near $\pi/2$, and it is under this condition that the polarization angle exerts a noticeable influence on the \ac{SNR}.
In contrast, at the poles of Figure \ref{fig:polar}, the contour lines are almost coincident with the lines of latitude, indicating that the influence of the polarization angle $\psi$ can be disregarded in this area.

\begin{figure}[htbp]
  \centering
  \includegraphics[width=0.4\textwidth]{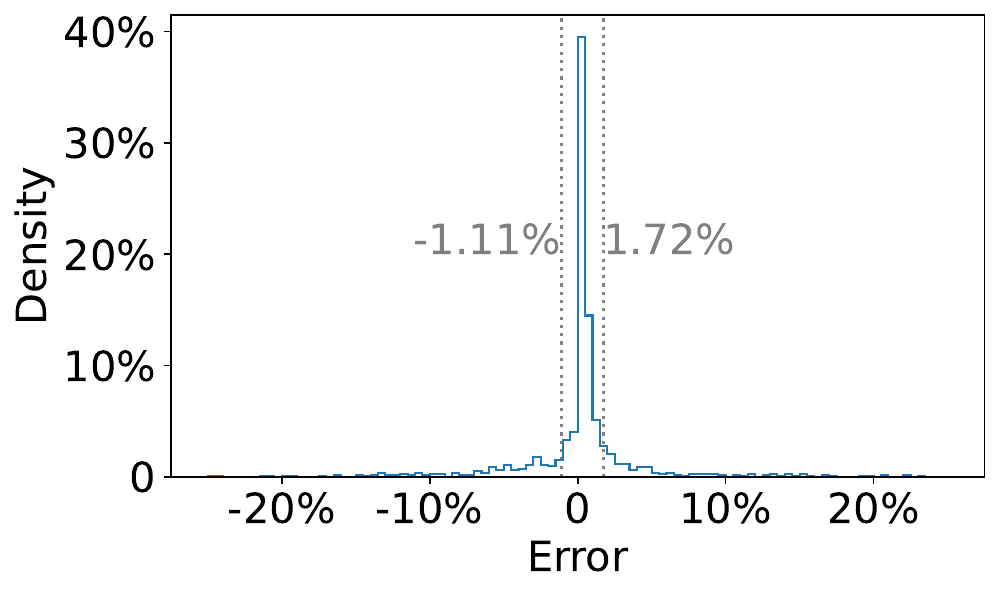}
  \caption{The density plot of the relative error for Equation \ref{equ:Pres_factorization}, calculated using 1,000 random scenarios. The $1\sigma$ error of the relative error is represented by a grey dotted line.}
  \label{fig:Pres_err}
\end{figure}

Although the response factor is challenging to decompose analytically into sky position and polar angle components, we suggest an alternative approach that entails the multiplication of the results derived from the pair of diagrams presented in Figure \ref{fig:ResponseFactor} to ascertain the response factor:
\begin{eqnarray}\label{equ:Pres_factorization}
P_{\rm res} = P(\lambda,\beta) \times P(\psi,\iota).
\end{eqnarray}
To verify the reliability of this calculation, we randomly sample 1,000 points to assess whether this approximation introduces any error. 
All sampled points fall within the detectable range of masses between $10^4 \sim 10^7 ~M_{\odot}$.
As depicted in Figure \ref{fig:Pres_err}, the $1\sigma$ relative error arising from this method remains well within an acceptable range of 2\%.

\section{Stellar-mass Black Holes}\label{sec:SBBH}

Besides \acp{MBHB}, we are further exploring the feasibility of deriving an analytical \ac{SNR} formula for inspiral \acp{SBBH}.
Due to the relatively high frequency at which \acp{SBBH} evolves within TianQin's sensitive band, the \ac{LFA} is no longer applicable. 
Conversely, the \ac{PSD} formula of TianQin in \ac{ASA} case can be simplified by \ac{SWA} as follows:
\begin{eqnarray}\label{equ:PSD_highFreq}
S_n = \frac{10}{3} L^2 S_p \times \left[1+0.6\left(\frac{2\pi fL}{c}\right)^2\right]
\end{eqnarray}
Using the same methodology of \ac{MBHB}, we can expand the \ac{SNR} integral and retain only the zero-order term.:
\begin{eqnarray}\label{equ:SBBH_result}
\fl \rho(T_{\rm obs}) &=& \frac{1.15829}{c^{11/4}\sqrt{N_p}} \left(G\mathcal{M}\right)^{5/4} \frac{L}{D_L} \times \left(\sqrt{t_c}-\sqrt{t_c-T_{\rm obs}}\right)^{1/2} \nonumber \\
    \fl &\approx& 8.4 \left(\frac{\mathcal{M}}{100~M_{\odot}}\right)^{5/4} \frac{100~{\rm Mpc}}{D_L} \times \left(\sqrt{\frac{t_c}{\rm 1~month}}-\sqrt{\frac{t_c-T_{\rm obs}}{\rm 1~month}}\right)^{1/2},
\end{eqnarray}
and the lowest-order correction:
\begin{eqnarray}\label{equ:SBBH_1o}
C &=& -\frac{0.388101}{c^{13/3}N_p} \left(G\mathcal{M}\right)^{5/3} \frac{L^{10/3}}{D_L^2} \left[\mathcal{F}(t_c)-\mathcal{F}(t_c-T_{\rm obs})\right], \nonumber \\
    &\approx& 162 \left(\frac{\mathcal{M}}{100~M_{\odot}}\right)^{5/3} \left(\frac{100~{\rm Mpc}}{D_L}\right)^2 \left[\mathcal{F}(t_c)-\mathcal{F}(t_c-T_{\rm obs})\right],
\end{eqnarray}
where,
\begin{eqnarray}
\mathcal{F}(t) &=& {\rm arctan}\left[0.57735-\frac{0.577948~c^{7/12} L^{2/3}}{\left(G\mathcal{M}\right)^{5/12} t^{1/4}}\right], \nonumber \\
    &\approx& {\rm arctan}\left[0.57735-\frac{0.23866}{\left(\frac{\mathcal{M}}{100~M_{\odot}}\right)^{5/12} \left(\frac{t}{\rm 1~month}\right)^{1/4}}\right]
\end{eqnarray}

\begin{figure*}[htbp]
  \centering
  \subfigure[w/o lowest-order correction.]{\includegraphics[width=0.45\textwidth]{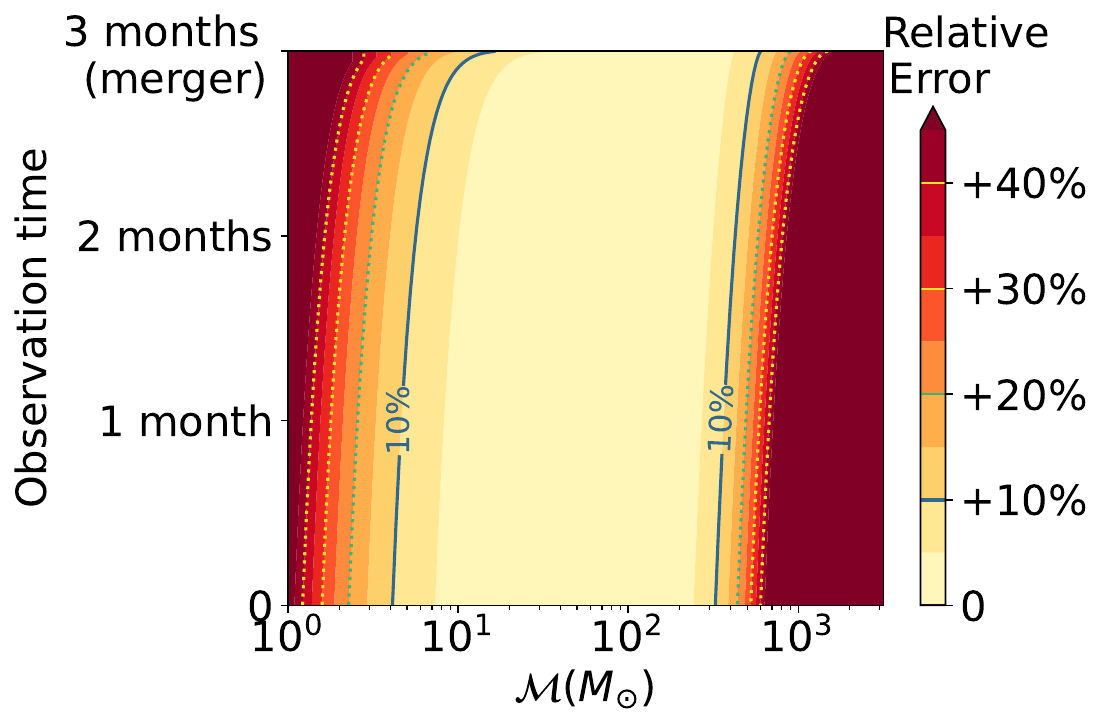}}\label{fig:SNR_appro_highF}
  \subfigure[with lowest-order correction.]{\includegraphics[width=0.45\textwidth]{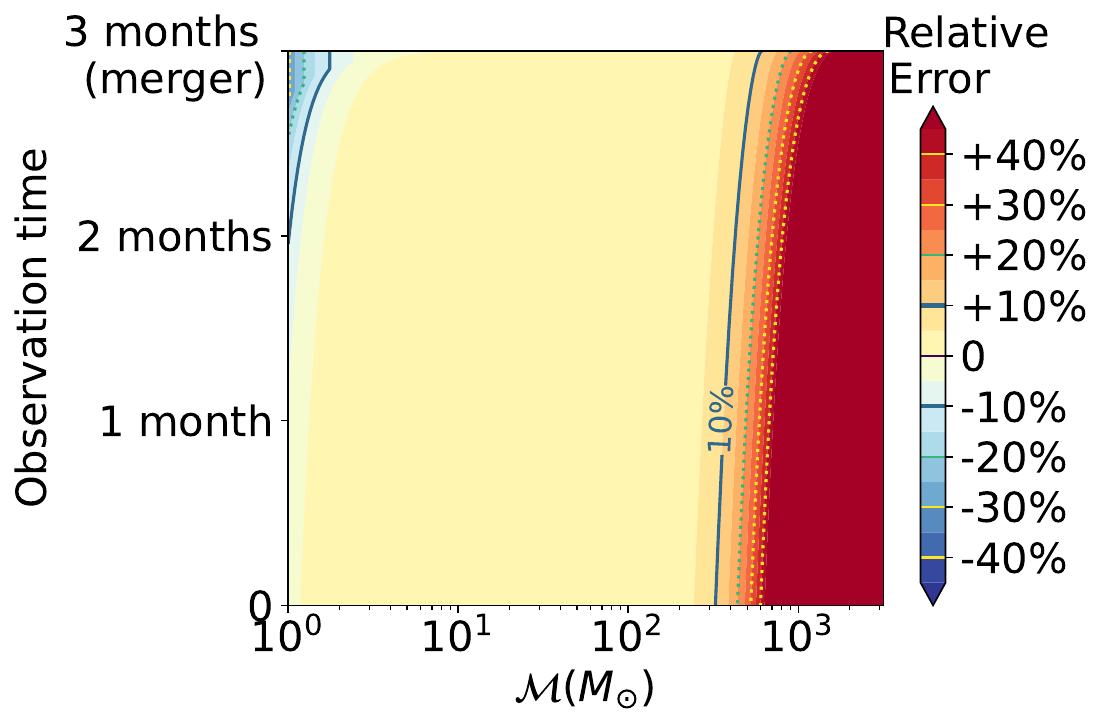}}\label{fig:SNR_appro_1_highF}
  \caption{The \ac{SNR} estimation error of \acp{SBBH} for TianQin varies with the chirp mass and observation time. The left panel uses the simplest zeroth-order \ac{SNR} formula for estimation, while the right panel considers the lowest-order correction.}
  \label{fig:snr_highF}
\end{figure*}

In our test, as shown in Figure \ref{fig:snr_highF}, we found that in the case of \ac{ASA}, our formula provided good estimation accuracy for \acp{SBBH}. 
Without incorporating the lowest-order correction, we could make good estimates of the \ac{SNR} for signals ranging from $7 \sim 400 ~ M_{\odot}$. 
The \ac{SNR} was overestimated for signals outside this range. 
When the lowest-order correction was applied, the \ac{SNR} of most signals with masses below $400 ~ M_{\odot}$ could be accurately calculated.

\begin{figure}[htbp]
  \centering
  \includegraphics[width=0.4\textwidth]{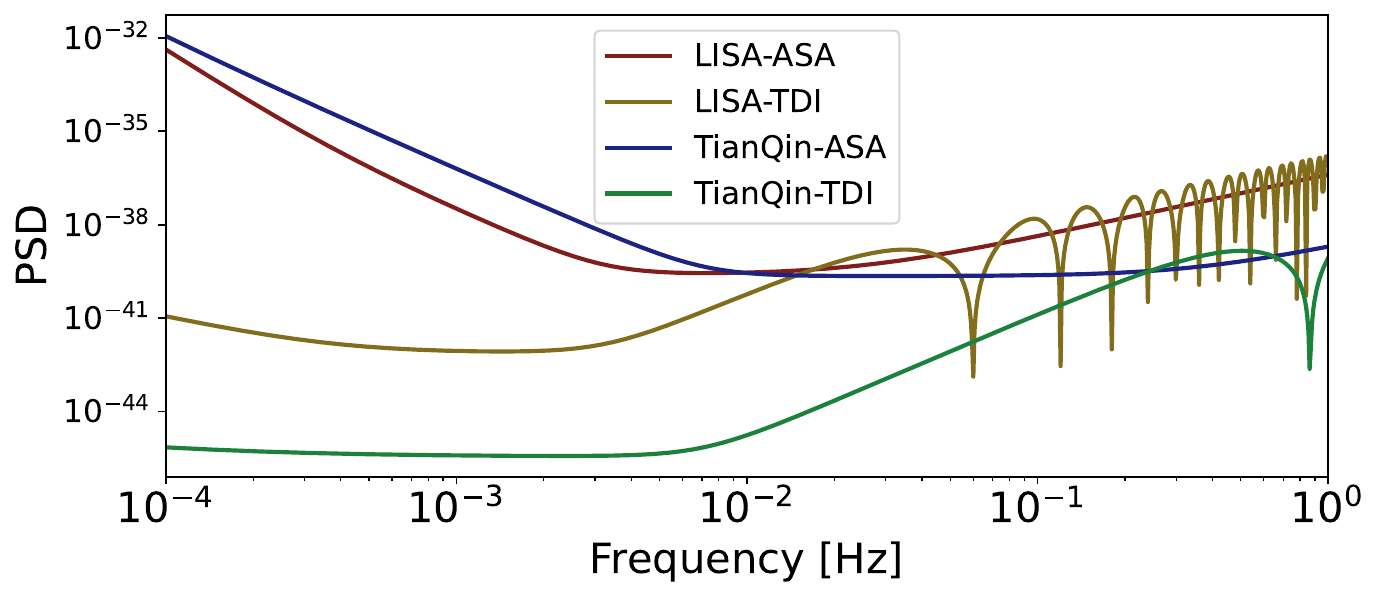}
  \caption{The \ac{PSD} curves of TianQin and \ac{LISA} under \ac{ASA} and \ac{TDI}-A/E channel response conditions.}
  \label{fig:PSD}
\end{figure}

Despite our formula providing good estimations in the case of \ac{ASA}, the \ac{SBBH} signal with response is challenging to compute analytically. 
As illustrated in Figure \ref{fig:PSD}, after considering the TDI response, the low-frequency component of the \ac{PSD} remains smooth, allowing the \ac{SNR} of the \ac{MBHB} signal with response to be analytically approximated using \ac{LFA}. 
While in the high-frequency regime where \ac{SBBH} resides, the \ac{PSD} becomes highly complex and lacks a satisfactory formula for approximation.
However, it is important to note that while we consider a three-month observation for all systems, \acp{SBBH} can keep inspiraling within the sensitive bands of space detectors for years. Therefore, our estimation of the \ac{ASA} \ac{SNR} will also be applicable for such long-duration inspiral-only observations.

\section{Conclusion}\label{sec:conculsion}

In this work, we present a simplified analytic formula for the \ac{SNR} of \ac{BBH} systems in TianQin.
The analytical formula for low-frequency scenarios is applicable to signals from \acp{MBHB} with chirp masses above approximately $2\times10^3~M_\odot$, while the analytical formula for high-frequency scenarios is suitable for signals from \acp{SBBH} with chirp masses below approximately $400~M_\odot$.
In the Table \ref{tab:summary}, we summarize all the formulas, their applicability, and the assumptions in this paper.

\begin{table}
\caption{\label{tab:summary}The table summarize all the formulas, their applicability, and the assumptions in this paper.}
\footnotesize
\begin{tabular}{@{}llcl}
\br
Sources & SNR formulae & Applicability($M_{\odot}$)$^1$ & Assumptions\\
\mr
\multirow{3}{*}{MBHB}
     & Zero-order formulae (\ref{equ:ASA_result}) & $2\times10^3 \sim 5\times10^4$ & LFA(\ref{equ:ASA_sensitivity_lowF}), $f$-independent $S_a$(\ref{equ:ASA_sensitivity_lowF})\\
     & Lowest-order formulae (\ref{equ:ASA_1o},\ref{equ:ASA_corr}) & $2\times10^3 \sim 2\times10^5$ & LFA(\ref{equ:ASA_sensitivity_lowF}), Integral expansion(\ref{equ:corr_expand}) \\
     & Response factor (\ref{equ:nonASA_result},\ref{equ:Pres_factorization})
     & $\geq 2\times10^3$
     & LFA(\ref{equ:LFA_yslr},\ref{equ:LFA_aet},\ref{equ:SAElowF}), $P_{\rm res}$ factorization \ref{equ:Pres_factorization}\\
\mr
\multirow{2}{*}{SBBH}
     & Zero-order formulae (\ref{equ:SBBH_result}) & $7 \sim 400$ &  SWA(\ref{equ:PSD_highFreq})\\
     & Lowest-order formulae (\ref{equ:SBBH_1o}) & $1 \sim 400$ & SWA(\ref{equ:PSD_highFreq}) \\
\br
\end{tabular}\\
$^1$The applicability indicates the mass range where formula error stays below 10\% one month prior to merger. Except near merger, the applicability shows little variation over time.
\end{table}

On the one hand, our findings can enable researchers in various fields to quickly and accurately estimate the \ac{SNR} of \ac{GW} signals from different types of \ac{BBH} systems. 
On the other hand, our results reveal the accumulation pattern of the \ac{SNR} for \ac{BBH} signals, which can be applied to the identification of future signals and the construction of detection statistics.

We evaluated the significance of errors resulting from using \ac{SNR} analytical formulae. 
For TianQin, our formula of \ac{MBHB} can accurately estimate the inspiral signals within the mass range of $2\times10^3 \sim 5\times10^4~M_{\odot}$. 
Incorporating a lowest-order correction can extend the range of accurate estimates between $2\times10^3 \sim 2\times10^5~M_{\odot}$.
However, for \ac{LISA}, achieving good estimation accuracy is challenging, regardless of whether the lowest-order correction is considered.

We derived an accurate analytical formula for the \ac{SNR} of \acp{SBBH} in the \ac{ASA} case.
The formula can accurately estimate the inspiral signals within the mass range of $7$ to $400~M_{\odot}$. 
Upon the application of the lowest-order correction, the \ac{SNR} of the majority of signals with chirp mass below $400 ~ M_{\odot}$ could be accurately calculated.
Due to the complexity of the \ac{PSD} at high frequencies, it remains challenging to provide a precise analytical formula for the \ac{SNR} of \ac{SBBH} signals with response.
We will delve into this scenario in the future.

Specifically, we computed the \ac{SNR} analytic formulae for both the \ac{ASA} and non-average conditions. 
The sole discrepancy between these two scenarios is a coefficient, which we have termed the "response factor".
This factor is exclusively affected by the sky position and the polar angle and does not vary over time. 
Despite the difficulty in performing an analytical calculation, we have developed a simple numerical approach to determine its value across different sky positions and polar angles. 
By sampling points over the parameter spaces including either the sky position or the polar angle and multiplying the evaluated values, we obtain an approximate estimation result with an error margin of within 2\%.

This paper considers a relatively ideal scenario. 
In the future, TianQin is likely to encounter issues such as data gaps. 
However, previous work\cite{LuWang2025} has shown that data gaps during the inspiral phase do not significantly change the \ac{SNR} of \acp{BBH}.
If necessary, adjusting the frequency integral boundaries for a piecewise calculation is straightforward.
The total \ac{SNR} can be represented as the square root of the sum of the squares of the \acp{SNR} of multiple data segments.

\section*{Acknowledgment}\label{sec:acknowledgment}

This work has been supported by the National Key Research and Development Program of China (No. 2023YFC2206700), and the Natural Science Foundation of China (Grants No. 12173104, No. 12261131504).
We thank Jian-Wei Mei and Jian-Dong Zhang for their helpful comments. 

\appendix
\section{Detectors Setting}\label{sec:det_set}

For TianQin\cite{Luo:2016CQGra, Li:2023arXiv230915020L}, the arm length $L = \sqrt{3} \times 10^8 ~{\rm m}$. 
The residual acceleration noise $S_a$ of the test masses in the satellites, and the position measurement noise $S_p$ can be represented as:
\begin{eqnarray}
S_a =& N_a \times \left(1+\frac{10^{-4}~Hz}{f}\right), \label{equ:SaTQ}\\
S_p =& N_p,
\end{eqnarray}
where, $N_a=10^{-30} ~{\rm m^2/s^4/ Hz}$, $N_p=10^{-24} ~{\rm m^2/ Hz}$.

For LISA\cite{LISA:2017arXiv, Li:2023arXiv230915020L}, the arm length $L = 2.5 \times 10^9 ~{\rm m}$. 
$S_a$ and $S_p$ can be represented as:
\begin{eqnarray}
S_a =& N_a \times \left[1+\left(\frac{0.4 \times 10^{-3} ~{\rm Hz}}{f}\right)^2\right] \times \left[1+\left(\frac{f}{8 \times 10^{-3} ~{\rm Hz}}\right)^4\right], \label{equ:SaLISA}\\
S_p =& N_p \times \left[1+\left(\frac{2 \times 10^{-3} ~{\rm Hz}}{f}\right)^4\right].
\end{eqnarray}
where, $N_a=9 \times 10^{-30} ~{\rm m^2/s^4/ Hz}$, $N_p=2.25 \times 10^{-22} ~{\rm m^2/ Hz}$.

Moreover, TianQin is only sensitive to signals at the frequency range of $f_{\rm low}=10^{-4}$ Hz to $f_{\rm high}=1$ Hz. 
Consequently, any signal outside this range should be truncated.
For low-frequency cutoff, we discard data before $t_l$:
\begin{eqnarray}
t_c' &= t_c - t_l, \quad T_{\rm obs}' = T_{\rm obs} - t_l,& {\rm if} ~t_l > 0, \\
    t_c' &= t_c, \quad T_{\rm obs}' = T_{\rm obs},& {\rm if} ~t_l \leq 0. 
\end{eqnarray}
For high-frequency cutoff, we discard data after $t_h$:
\begin{eqnarray}
T_{\rm obs}' = {\rm min} \{T_{\rm obs}, t_h \},
\end{eqnarray}
where, the $t_l$ and $t_h$ can be calculated from Equation \ref{equ:ft_relation}:
\begin{eqnarray}
t_l &= t_c - \frac{5}{(8\pi f_{\rm low})^{8/3}}\left(\frac{G\mathcal{M}}{c^3}\right)^{-5/3} \approx t_c - 18 ~{\rm h} \times \left(\frac{10^7 ~M_{\odot}}{\mathcal{M}}\right)^{5/3}. \\
t_h &= t_c - \frac{5}{(8\pi f_{\rm high})^{8/3}}\left(\frac{G\mathcal{M}}{c^3}\right)^{-5/3} \approx t_c - 180 ~{\rm h} \times \left(\frac{1 ~M_{\odot}}{\mathcal{M}}\right)^{5/3}.
\end{eqnarray}

\section{ASA SNR with low-frequency approximation}\label{sec:cal_asa}

The \acp{GW} emitted by \ac{BBH} systems can be described by multiple modes, each corresponding to different oscillation patterns as the waves propagate through spacetime.
These modes are often classified by their angular harmonics, denoted by the symbol $(l,m)$ representing the multipole moment of the wave.
The quadrupole 22 mode, with $l = m = 2$, is the most significant and typically the only feature in the \ac{GW} signal observed at large distances from the source. 
Therefore, in this work, we only consider \acp{GW} produced by the 22 mode. 
Consequently, in \ac{ASA} condition, the coefficients of the 22 mode \cite{Poisson:1993vp} will be retained:
\begin{eqnarray}\label{equ:ASA_amplitude}
\mathcal{A}_{ASA}(f) = \sqrt{\frac{5}{16\pi}} \times \mathcal{A}(f).
\end{eqnarray}

Additionally, the \ac{ASA} sensitivity curve of TianQin can be represented as \cite{Li:2023arXiv230915020L}:
\begin{eqnarray}\label{equ:ASA_sensitivity}
S_n = \frac{10}{3} \frac{1}{L^2} \left[\frac{4S_a}{(2\pi f)^4}+S_p\right] \times \left[1 + 0.6\left(\frac{f}{f_{\ast}}\right)^2\right],
\end{eqnarray}
where $f_{\ast} \equiv c/(2\pi L) \approx 0.28 ~{\rm Hz}$. 

Given that the inspiral signals from \acp{MBHB} are generally found in the low-frequency spectrum with $f\ll f_\ast$
, the $S_p$ and $\left[1 + 0.6(f/f_{\ast})^2\right]$ terms of the Equation \ref{equ:ASA_sensitivity}, which primarily influence high frequencies, can be omitted.
Besides, we can firstly assume that the residual acceleration noise $S_a=N_a$ is frequency-independent, and then the sensitivity curve of TianQin can be approximated as:
\begin{eqnarray}\label{equ:ASA_sensitivity_lowF}
S_n \simeq \frac{5}{6\pi^4} \frac{1}{L^2} N_a f^{-4}
\end{eqnarray}

By combining the amplitude (Equation \ref{equ:ASA_amplitude}) and sensitivity curve (Equation \ref{equ:ASA_sensitivity_lowF}) under the \ac{ASA} condition, we can calculate the square of the \ac{SNR}:
\begin{eqnarray}  
\rho^2 &= 4 \int^{f_{\rm max}}_{f_{\rm min}} {\frac{\mathcal{A}_{ASA}^2(f)}{S_n(f)}} {\rm d}f = \frac{\pi^{8/3}}{c^3} \left(\frac{L}{D_L}\right)^2 \frac{1}{N_a} \left(G\mathcal{M}\right)^{5/3} \int^{f_{\rm max}}_{f_{\rm min}} f^{5/3} {\rm d}f, \nonumber \\
    &= \frac{15 c^2}{2048 N_a t_c} \left(\frac{L}{D_L}\right)^2 \frac{T_{\rm obs}}{t_c-T_{\rm obs}},
\end{eqnarray}
It is of particular interest to observe that within this \ac{SNR} formula, all elements associated with mass, originating from both the time-frequency relation and the amplitude, have been effectively eliminated.

By extracting the square root of the aforementioned equation, we derive the relationship that governs the accumulation of \ac{SNR} over time:
\begin{eqnarray}
\rho (T_{\rm obs}) &= \sqrt{\frac{15}{2048}} \frac{L}{D_L} \frac{c}{\sqrt{N_a t_c}} \times \sqrt{\frac{t_{obs}}{t_c-t_{obs}}}, \nonumber \\
    &\approx 185.2 \times \frac{1~{\rm Gpc}}{D_L} \sqrt{\frac{1~{\rm week}}{t_c}} \times \sqrt{\frac{T_{\rm obs}}{t_c-T_{\rm obs}}}.
\end{eqnarray}

\section{Calculation of the correction}\label{sec:cal_corr}

The error in our formulas within the low frequency region of \acp{MBHB} mainly comes from the frequency dependence of acceleration noise, as indicated in Equation \ref{equ:SaTQ} and \ref{equ:SaLISA}. 
Considering the frequency dependence, the SNR squared becomes:
\begin{eqnarray}\label{equ:corr_expand}
\rho_{\rm corr}^2 (t) \approx \frac{\mathcal{M}^{5/3}}{N_a \times D_L {}^2} \times \sum_{i=0}^{n} c_i \left[x(t)-x(0)\right]^{i-8/3},
\end{eqnarray}
where $c_i$ represents distinct constants, and $x\equiv \mathcal{M}^{5/8}\left(t_c-t\right)^{3/8}$.
It should be noted that for LISA, $c_1 = 0$, so its lowest-order correction is $i = 2$.

\section{Calculation of the response factor}\label{sec:cal_non_asa}

The response of a space-based \ac{GW} detector can be characterized by the single-link observables, represented as $y_{slr}=(\nu_r-\nu_s)/\nu$, which quantifies the relative laser frequency shift between the transmitting spacecraft (s) and the receiving spacecraft (r) along the link (l) \cite{Li:2023arXiv230915020L, Vallisneri:2005PhRvD, Marsat:2021PhRvD}. 
The relationship between the observable and the source waveform is described by:
\begin{eqnarray}
\tilde{y}_{slr} = G^{22}_{slr}(f,t) \times \mathcal{A} e^{{\rm i} \Phi},
\end{eqnarray}
where, $\Phi$ is the waveform phase, and $G^{22}_{slr}(f,t)$ denotes the transfer function \cite{Li:2023arXiv230915020L, Marsat:2021PhRvD}:
\begin{eqnarray}
G^{22}_{slr}(f,t) =& -\frac{{\rm i}\pi fL}{2c} \times {\rm sinc} \left[ \frac{\pi fL}{c} (1- k\cdot n_l) \right ] \nonumber \\
    & \times \exp \left [{\rm i}\pi f \left( \frac{L+ k \cdot (p_{r}+p_{s})}{c} \right) \right] \times P_{slr}(t),
\end{eqnarray}
with $k$ representing the wave propagation vector, and $s, r$ denoting the transmitting and receiving spacecraft, respectively. 
The expression $P_{slr}(t) = n_l(t) \cdot P^{22} \cdot n_l(t)$ defines the inner product of the 2-2 mode polarization tensor $P^{22}$ and the link unit vectors $n_l$.

By eliminating the phase terms that do not affect the \ac{SNR} and employing a \ac{LFA}, the amplitude of the observable can be simplified as:
\begin{eqnarray}\label{equ:LFA_yslr}
\tilde{y}_{slr} \simeq -\frac{{\rm i}\pi fL}{2c} \times P_{slr}(t) \times \mathcal{A},
\end{eqnarray}

Here we employ the standard set of orthogonal \ac{TDI} observables, namely A, E, and T.
The waveforms in these channels can be expressed as \cite{Li:2023arXiv230915020L, Marsat:2021PhRvD}:
\begin{eqnarray}
\tilde{A}, \tilde{E} & = {\rm i} \sqrt{2} \sin\left(\frac{f}{f_{\ast}}\right) \exp\left({\rm i}\frac{f}{f_{\ast}}\right) \times \tilde{a}, \tilde{e}, \\
\tilde{T} & = 2 \sqrt{2} \sin\left(\frac{f}{f_{\ast}}\right) \sin\left(\frac{f}{2f_{\ast}}\right) \exp\left({\rm i}\frac{3f}{2f_{\ast}}\right) \times \tilde{t},
\end{eqnarray}
where the $\tilde{a}$, $\tilde{e}$ and $\tilde{t}$ terms can be simplified, under the \ac{LFA}, to \cite{Marsat:2021PhRvD}:
\begin{eqnarray}\label{equ:LFA_aet}
\tilde{a} \simeq & \, 4\tilde{y}_{31} - 2\tilde{y}_{23} - 2\tilde{y}_{12}, \\
\tilde{e} \simeq & \, 2\sqrt{3} \left[\tilde{y}_{12} - \tilde{y}_{23} \right], \\
\tilde{t} \simeq & \, 0.
\end{eqnarray}

Therefore, the absence of signals in the T-channel data does not contribute to the \ac{SNR}. 
Meanwhile, the signal in the A and E channels can be further simplified as:
\begin{eqnarray}
\tilde{A} \simeq & \, 2\sqrt{2} \mathcal{A} \left(\frac{\pi fL}{c}\right)^2 \left(2P_{31} - P_{23} - P_{12}\right), \label{equ:Asignal} \\
\tilde{E} \simeq & \, 2\sqrt{6} \mathcal{A} \left(\frac{\pi fL}{c}\right)^2 \left(P_{12} - P_{23}\right). \label{equ:Esignal} 
\end{eqnarray}
Additionally, the noise \ac{PSD} in the A, E channels can be represented as \cite{Li:2023arXiv230915020L}:
\begin{eqnarray}
\fl S_{A} &=& S_{E} \nonumber \\
    \fl &=& 8 \sin^2\left(\frac{f}{f_{\ast}}\right) \left[4 \left(1 + \cos \left(\frac{f}{f_{\ast}}\right) + \cos \left(\frac{2f}{f_{\ast}}\right)\right)S_{\rm acc} + \left(2 + \cos \left(\frac{f}{f_{\ast}}\right)\right)S_{\rm oms}\right],
\end{eqnarray}
where $S_{\rm acc} = S_a \left(1/(2\pi f c)\right)^2$ is the acceleration noise, primarily dominant at low frequencies. And $S_{\rm oms} = S_p \left((2\pi f)/c\right)^2$ is the displacement or position noise, primarily dominant at high frequencies.
Therefore, after also applying the \ac{LFA} and the frequency-independent $S_a$ approximation, the low-frequency \ac{TDI} noise \ac{PSD} of TianQin can be simplified to:
\begin{eqnarray}\label{equ:SAElowF}
S_{A} =S_E \simeq \frac{96 ~ S_a L^2}{c^4} \simeq \frac{96 ~ N_a L^2}{c^4}.
\end{eqnarray}

Ultimately, using the signals (Equation \ref{equ:Asignal},(\ref{equ:Esignal})) and noise \ac{PSD} (Equation \ref{equ:SAElowF}) in the A and E channels, we can calculate the \ac{SNR} for \acp{MBHB} in non-average scenarios:
\begin{eqnarray}
\rho^2 &= 4 \int^{f_{\rm max}}_{f_{\rm min}}{\frac{\tilde{A}^2(f)+\tilde{E}^2(f)}{S_{A}}} {\rm d}f, \nonumber \\
& \simeq \frac{\pi^{8/3}}{2 c^3} \left(\frac{L}{D_L}\right)^2 \frac{1}{N_a} \left(G\mathcal{M}\right)^{5/3} \int^{f_{\rm max}}_{f_{\rm min}} P_{\rm res}^2 f^{5/3} {\rm d}f,
\end{eqnarray}
where, the response factor $P_{\rm res}$ is defined as:
\begin{eqnarray}
P_{\rm res}^2 = \frac{8\pi}{9} \times \left(P_{12}^2 + P_{23}^2 + P_{31}^2 - P_{12}P_{31} - P_{23}P_{31} - P_{12}P_{23}\right).
\end{eqnarray}

It is noteworthy that, although each $P_{slr}$ fluctuates with time and frequency, the aggregation into $P_{\rm res}$ form renders it temporally invariant, with variation contingent only upon the sources' sky positions and polar angles.
This enables the extrication of this term from the integral, thereby maintaining the simplicity of the \ac{SNR} formulation.
Consequently, the \ac{SNR} formula in the non-average form simply incorporates an extra $P_{\rm res}$ term compared to the \ac{ASA} form (Equation \ref{equ:ASA_result}):
\begin{eqnarray}
\rho (T_{\rm obs}) \simeq \sqrt{\frac{15}{2048}} \frac{L}{D_L} \frac{c}{\sqrt{N_a t_c}} P_{\rm res}\times \sqrt{\frac{T_{\rm obs}}{t_c-T_{\rm obs}}},
\end{eqnarray}

\section*{References}
\bibliographystyle{iopart-num}
\bibliography{iopart-num}

\end{document}